\documentclass[preprint2]{aastex}

\usepackage{comment}
\usepackage{graphicx}

\usepackage{mathtools}

\usepackage{natbib}
\usepackage{float}

\usepackage{booktabs}

\usepackage{wrapfig}
\usepackage{ulem}

\usepackage{xcolor}

\newcommand{\Lsun}{$\rm L_{\odot}$}
\newcommand{\Msun}{$\rm M_{\odot}$}
\newcommand{\Rsun}{$\rm R_{\odot}$}
\newcommand{\Mdot}{$\rm \dot{M}$}
\newcommand{\Av}{A$_V$}
\newcommand{\msunyr}{$\rm{M_{\sun} \, yr^{-1}}$}
\newcommand{\Teff}{$\rm T_{eff}$}
\newcommand{\mic}{$\rm \mu$m}

\shortauthors{Rubinstein et al.}

\begin{document}
%\captionsetup[figure]{labelfont={bf},name={Fig.},labelsep=space}

\title{A Cavity of Large Grains in the Disk Around the Group II Herbig Ae/Be Star HD 142666}
%Examining HD 142666's Disk: A Group II Object With an Inner Cavity of Large Grains
%A Cavity of Large Grains in the Disk Around the Group II Source HD 142666
%A Cavity in the Distribution of Large Grains for the Disk Around the Group II Source HD 142666
%The Group II Source HD 142666 and its Cavity of Large Grains

\author{A. E. Rubinstein\altaffilmark{1}, E. Mac\'ias\altaffilmark{2}, C. C. Espaillat\altaffilmark{2}, K. Zhang\altaffilmark{1}, N. Calvet\altaffilmark{1}, C. Robinson\altaffilmark{2}}

\affil{\footnotesize \textsuperscript{1} Department of Astronomy, University of Michigan, Ann Arbor, MI 48109, USA}
\affil{\footnotesize \textsuperscript{2} Department of Astronomy, Boston University, 725 Commonwealth Avenue, Boston, MA 02215, USA}

\begin{abstract}
Herbig Ae/Be (HAeBe) stars have been classified into Group I or Group II, which were initially thought to be flared and flat disks, respectively. Several Group I sources have been shown to have large gaps, suggesting ongoing planet formation, while no large gaps have been found in the disks of Group II sources. We analyzed the disk around the Group II source, HD 142666, using irradiated accretion disk modeling of the broad-band spectral energy distribution along with the 1.3 millimeter spatial brightness distribution traced by Atacama Large Millimeter and Submillimeter Array (ALMA) observations. Our model reproduces the available data, predicting a high degree of dust settling in the disk, which is consistent with the Group II classification of HD 142666. 
In addition, the observed visibilities and synthesized image could only be reproduced when 
including a depletion of large grains out to $\sim$16 au in our disk model, although the ALMA observations did not have enough angular resolution to fully resolve the inner parts of the disk.
These results may suggest that some disks around Group II HAeBe stars have cavities of large grains as well. Further ALMA observations of Group II sources are needed to discern how commonly cavities occur in this class of objects, as well as to reveal their possible origins. 
\end{abstract}


\section{Introduction} %Sect 1
\label{intro}

Protoplanetary disks have long been thought to be the birthplaces
of planetary systems. Early models of dust
evolution predicted that the
dust inherited from the ISM grew by 
colliding, sticking together,
and settling towards the disk
midplane, where further 
interactions made them grow towards
planetesimals and planets
\citep{weidenschilling97}.
Observations have supported and expanded this
view; in particular, they have revealed
disk structures that have been
attributed to interactions between
the disk and forming planets
\citep[cf.][]{espaillat14}.

{Before the advent of
high-spatial resolution instruments,
disk properties were often inferred
from studies of the spectral energy distributions
(SEDs) of young objects.
In particular, 
such studies have focused on young,
intermediate-mass 
stars, known as Herbig Ae/Be stars (HAeBe) \citep{waters98}, 
which have
been separated into two groups,
Group I and Group II. 
%This classification
%is based on differences derived
%from observations in the mid-IR range from which the
%state of the dust in the system has been inferred.
%Systematic differences in other
%wavelength ranges have been sought 
%to further understand the nature of the
%two classes.
This classification came from
fitting the mid-IR (15 -- 45 \mic)
portion of HAeBe stars' SEDs.
The SEDs of Group I stars
required both a blackbody and power law to fit, while those of Group
II stars could be fit with only a power law \citep{meeus01}.
On average, it was noted that Group I stars have larger IR excesses when compared with Group II stars
\citep{dominik03}, which has been quantified
by using IRAS \citep{vanboekel03}
and Herschel data \citep{pascual16}. 
Specifically, the latter study 
found that Group II
sources have a lower ratio of far-IR (20 -- 200 \mic) to near-IR (2 -- 5 \mic) flux
than do Group I sources.
In turn, these differences have been interpreted as Group I
objects having ``flared'' disks and Group II objects having ``flat''
disks. Flared disks are defined such that
the height of the disk increases rapidly with radius,
capturing more stellar light,
and therefore being hotter \citep{kenyon87} 
than flat disks. 
In fact, flared, Group I sources were thought to evolve
into flatter, Group II sources due to dust growth and settling \citep{dullemond04}.
%Dust growth and settling in the disk
%would then decrease
%the abundance of small grains in the upper
%disk layers, resulting in lower opacity to
%stellar radiation, cooler disks, and flatter
%SEDs 
%\citep{dullemond04,d'alessio06}.
%As a result, Group II objects were
%thought of as more evolved than Group I objects.

However, further 
differences were found that brought into question the evolutionary relationship between the groups.
SED modeling ($\sim$0.1 -- 1000 \mic), together with {N-band,
  Q-band, and 25 \mic} imaging, revealed that many Group I sources 
have large, \textgreater 20 au, inner gaps or cavities
\citep[e.g.,][]{maaskant13, honda15}. In contrast, based on
  N-band interferometry, Group II sources have only showed evidence of
  smaller, \textless1 au gaps close to the dust sublimation radius
  \citep[e.g. HD 142666, HD 144432;][]{menu15}.

The Atacama Large Millimeter and Submillimeter Array (ALMA) has provided images of protoplanetary disks with extreme detail, revealing substructures, such as narrow gaps and rings, that could not previously be studied \citep{alma2015, andrews16}. For Group II sources, prior ALMA observations have often focused on the disk's chemical content, gas dynamics, or disk asymmetries instead of substructures like gaps or cavities \citep[][]{guzman15, czekala15, white18, kataoka16}. Recently, though, ALMA observations revealed shallow, \textgreater20 au gaps around the Group II source HD 163296 \citep{isella16, zhang16, muroarena18}, suggesting the possible presence of gaps or cavities in other Group II sources. 

Here, we study the Group II 
%disk \citep{meeus01} of the 
HAe star HD 142666 
%(V1026 Sco)
\citep[V1026 Sco,][]{meeus01},
located in the Upper Sco OB-2 association (\citealt{sandell11}; d=150.1 pc; \citealt{gaia16a}).
%We studied the Group II disk \citep{meeus01} of the HAe star HD 142666
%(V1026 Sco), which is in the Upper Sco OB-2 association
%(\citealt{sandell11}; d=150.1 pc; \citealt{gaia16a}). 
%Using GAIA
%parallax measurements, the distance to HD 142666 is 150.1 pc
%\citep{gaia16a}.  
HD 142666 has been observed from near-IR wavelengths,
detecting the scattered light from the disk \citep{garufi17}, to
sub-mm wavelengths, detecting the unresolved emission from the
${}^{12}$CO J = 2 -- 1 \citep{panic09} and ${}^{12}$CO J = 3 -- 2
\citep{dent05} rotational lines. 

Some structure has also been found in the disk of HD 142666.
From SED modeling and interferometric
observations in the 1.6 --
  2.5 \mic \, and 8 -- 13 {\mic} wavelength ranges, 
\citet{schegerer13} inferred the existence of
a hole in the disk inside 0.3 au and a dust-free gap between 0.35 au and 0.8 au.
On the other hand,
\citet{banzatti18}
found evidence for no or a very small
cavity for this disk in their study combining near-IR (1.2 -- 4.5 \mic)
and CO rovibrational emission.
%
%
%We studied the Group II 
%disk \citep{meeus01} of the 
%HAe star HD 142666 
%(V1026 Sco)
%\citep[V1026 Sco,][]{meeus01},
% located in the Upper Sco OB-2 association (\citealt{sandell11}; d=150.1 pc; \citealt{gaia16a}). 
%Using GAIA parallax measurements, the distance to HD 142666 is 150.1 pc \citep{gaia16a}. 
%HD 142666 has been observed from NIR wavelengths, detecting the scattered light from the disk \citep{garufi17}, to sub-mm wavelengths, detecting the unresolved emission from the ${}^{12}$CO J = 2 -- 1 \citep{panic09} and ${}^{12}$CO J = 3 -- 2 \citep{dent05} rotational lines. 
%From 
%SED modeling in the 0.1 to 100 {\mic} range, as well as NIR and MIR interferometric measurements \textbf{
%in the 1.6 -- 2.5 \mic \ and 8 -- 13 {\mic}}
%wavelength ranges indicated
%that the disk 
%was shown 
%\textbf{to have a small, 
%\textbf{had a small, 
%\textless $\sim$1 
%$\le$ au gap starting at sub-au radii} \citep{schegerer13, vural14}. \textbf{While HD 142666 also has indications of a small inner cavity based on its CO radius and NIR excess \citep{banzatti18}, } 
So far, no evidence has 
suggested the presence of a large gap or cavity around
HD 142666 comparable to
those found in Group I sources.

In this paper, we present a physically self-consistent model of the SED and ALMA observations for the disk around HD 142666. In Section 2.1, we describe the archival photometry and spectra used for the SED modeling. In Section 2.2, we present archival 1.3 mm (Band 6) ALMA observations, and we show the results of an initial analytical fit to the observations. In Section 3.1, we describe the model and physical parameters used to reproduce the observations. In Section 3.2, we summarize the results, including a cavity extending radially out to $\sim$16 au in the disk's large dust grain distribution. Finally, in Section 4, we put this large cavity in context with relevant results involving cavities and gaps in HAeBe disks.

%define PTD

\section{Observations and Results} 

    \subsection{Stellar Properties and the SED}
To construct the SED of HD 142666 and its disk (Figure \ref{fig_res}), we gathered the IRS spectrum \citep{keller08} and visible to mm photometry from 2MASS, AKARI, IRAS, Tycho-2, WISE, and 
%the following references:
\citet{vanderveen94, sylvester96, malfait98, natta04, meeus12, mendigutia12}
and \citet{pascual16}. We note that we did not include photometry that overlapped with the IRS wavelengths (e.g. From AKARI or WISE) in order to avoid the wide bandpasses of these filters. 

The data was dereddened using the \citet{mathis90} reddening law  by first finding the extinction E(B-V) from the observed B-V color and the intrinsic (B-V)$_{0}$ color from \citet{k&h95}. Assuming R$_{V}$ = 3.1, we obtained A$_{V}$ = 0.992. 
The stellar luminosity was then derived using the dereddened V-band magnitude. This luminosity, together with the effective temperature, was compared with evolutionary tracks \citep{siess00} to obtain a stellar mass. These stellar properties are shown in Table \ref{tab_star}. 

The mass accretion rate (\Mdot) for HD 142666 has been measured to be between 1 x ${10}^{-8}$ and 2 x ${10}^{-7}$ \msunyr \citep{salyk13, mendigutia11, donehew11, garcialopez06}. Even though the range in these measurements could be due to intrinsic variability, these papers used different methods, making comparison difficult.

{
%\onecolumn
\begin{figure}[ht]
    \centering
    \includegraphics[width=\linewidth, trim={0.15cm 0.3cm 0.3cm 0.15cm},clip]{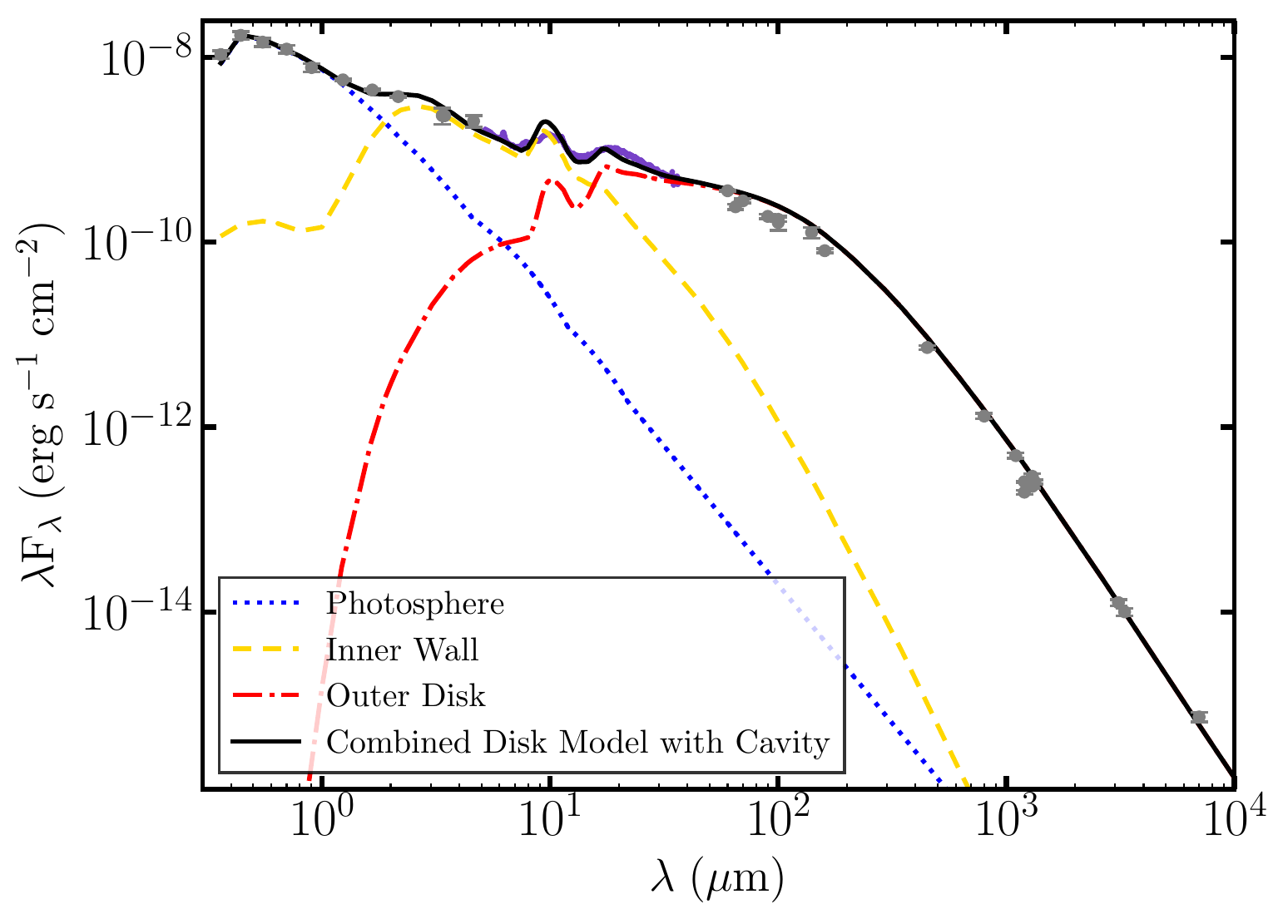}
    \caption{The SED including photometric data (gray dots) and the IRS spectrum (purple dots). The black curve shows the disk model including a cavity that fits the observations. Other curves represent contributions (i.e. from the star or disk) that are added to produce this model.}
\label{fig_res}
\end{figure}
}

\subsection{ALMA Observations}
    \label{obs}
We used Cycle 2 ALMA archival observations (PI: L. P\'erez) of the protoplanetary disk of HD 142666. The data were calibrated using the reduction package Common Astronomy Software Applications \citep[CASA:][]{casa07}. HD 142666 was observed on July 21, 2015 at Band 6 using an array of 44 antennas. The observations were flux calibrated with the QSOs J1517-2422 and J1627-2426 and bandpass calibrated with Titan. The total on-source time was 59.04 minutes. Two spectral windows (SPW) centered at 216.974 and 232.349 GHz were used to detect the continuum, each with a bandwidth of 1.875 GHz. 
%We note that the ALMA continuum observations are split into two SPWs that are separated by $\sim15$ GHz but have a high SNR individually. Thus, we have analyzed both SPWs separately and only show in the following figures the results for the SPW centered at 216.974 GHz.%The data were taken in 5 spectral windows (SPWs). Two SPWs were used to detect the continuum, each with a bandwidth of 1.875 GHz. Three other SPWs were tuned to detect the spectral lines XX (231.181 GHz), XX (219.246 GHz), and XX (218.778 GHz), each with a bandwidth of 0.468750 GHz.

We used the CLEAN task in CASA interactively to obtain deconvolved images from the observed visibilities. We chose a uniform weighting, resulting in a synthesized beam of $0\rlap."200\times0\rlap."165$ and an rms sensitivity of 0.22 mJy beam$^{-1}$. The lower left panel of Figure \ref{fig_img_obs} shows the resulting 1.38 mm continuum image. %We detect and resolve the dust emission of the protoplanetary disk around HD 142666. 
The continuum emission shows a resolved, compact disk without apparent evidence of non-axisymmetry or a gap.

{
%\onecolumn
\begin{figure*}[!ht]
    %\topfraction
    \centering
    \includegraphics[width=1\linewidth]{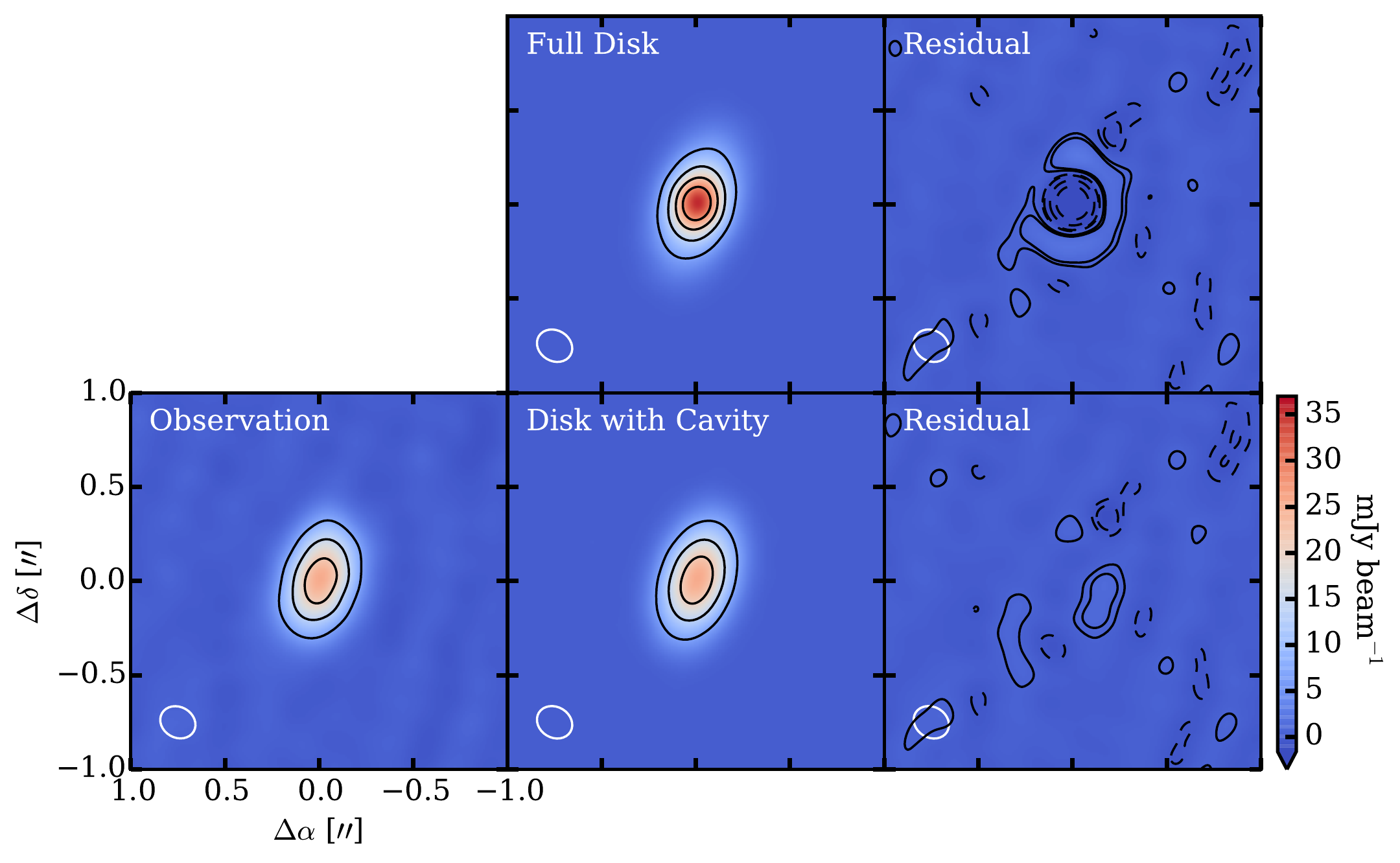}
    %\centering
    %\captionsetup{width=\textwidth}
    \caption{The synthesized ALMA continuum image of HD 142666 at 1.38 mm (bottom left), along with the simulated model image from a full disk (center, top) and a cavity-containing disk (center, bottom), and the residual images produced from subtracting the observed and model visibilities (right of the respective simulated image). The ellipse in the lower left corner shows the beamsize. The contours for the observation and cavity-containing disk model are 0.2, 0.4, 0.6, 0.8 times the maximum brightness of the image. The contours for the residual are $\pm$3, $\pm$5, $\pm$15, $\pm$30 times the rms of the image, 170 $\mu$Jy/beam.}
\label{fig_img_obs}
\end{figure*}
}

Enclosing the source inside a 1\arcsec$\times$1\arcsec \, box, we estimated a flux density of 113 $\pm$ 6 mJy, consistent with the results of a Gaussian fit to the image. The Gaussian fit was also used to estimate the central coordinates, position angle (PA; 161$^\circ$), FWHM (30 au), and inclination ($\sim$ 60$^\circ$) of the disk.  

Finally, we changed the phase center and deprojected the visibilities using small shifts in position, as well as different values of inclination and PA. The values that resulted in the minimum scatter in the real and imaginary parts of the visibilities were consistent with the results found by the Gaussian fit. The measured inclination agrees with previous studies based on H-band interferometric observations \citep{lazareff17} and modeling of mid-IR interferometry \citep{vural14}.

In Figure \ref{fig:fig_finalrad}, we show the deprojected, real part of the visibilities. We note that the ALMA continuum observations are split into two SPWs that are separated by $\sim15$ GHz but have a high SNR individually. Thus, we have analyzed both SPWs separately and only show the results for the SPW centered at 216.974 GHz (1.38 mm). The visibilities show a very steep decrease followed by a null at $\sim500$ k$\lambda$, suggesting the presence of a cavity in the disk \citep{hughes07}. To characterize the disk's intensity profile, we fit the visibilities using a parametric model of the surface brightness profile of the disk as in \citet{zhang16}. The top panels of Figure \ref{fig:fig_finalrad} show the results of this modeling. The resulting intensity profile shows a 
%deficit 
% "deficit" needs something to compare with
drop
in intensity at R\textless9.5 au. We used this intensity profile as a first estimate for the following physical disk modeling.

%\startlongtable
\begin{deluxetable}{ccccccc}
% \rotate
%\tabletypesize{\scriptsize}
\tablewidth{0pt}
\tablecaption{Stellar Properties}
\tablehead{
\colhead{Sp. Type} & \colhead{\Teff} & \colhead{\Av} & \colhead{L} & \colhead{M} & \colhead{R} \\
\colhead{} & \colhead{(K)} & \colhead{} & \colhead{(\Lsun)} & \colhead{(\Msun)} & \colhead{(\Rsun)}
}
\startdata
  A8 & 7580 & 0.992 & 14.3 & 2.0 & 2.2\\
\enddata
\tablecomments{We adopt a spectral type of A8 spectral type from \citet{mora01} and a {\Teff} of 7580 K following \citet{k&h95}. 
}
\label{tab_star}
\end{deluxetable}
% \hline\end{tabular}
%\clearpage

%\twocolumn
\section{Modeling} %Sect 3
\label{analysis}  
    
    \subsection{Modeling Procedure} 
    \label{diskmod}
        
We modeled the SED and ALMA continuum observations of HD 142666 using the physically self-consistent disk model by \citet{d'alessio06}. The model computes the radial and vertical structure of an irradiated accretion disk while also enforcing hydrostatic equilibrium. It includes a wall at the inner edge of the disk that is directly irradiated by the star. In addition, we added a tapered edge to the disk's surface density profile following $\rm {e}^{-({R/{R}_{d}})^{\gamma}}$, where $\rm {R}_{d}$ is the outer disk radius at which the tapering begins, and $\rm \gamma$ determines its steepness.
The viscosity is described by the
parameter $\alpha$ following
\citet{shakura&sunyaev73}.
We left it fixed because 
of its degeneracy with \Mdot \, \citep{d'alessio06}.
The dust in the disk was described by two populations
of grains with size distributions
$n(a) \propto a^{-3.5}$ between
a minimum size, $a_{min} = 0.005$ \, \mic,
and maximum size, $a_{max}$, which was kept
as a free parameter. The large grains (a$_{max}^{mid}$)
were concentrated at the midplane of the disk,
while the small grains (a$_{max}^{atm}$) were located
mostly in the upper layers
\citep{d'alessio06}. To include dust
settling, the population of small grains in the disk atmosphere
was depleted by a factor
$\epsilon$, which is the ratio of
the dust-to-gas mass ratio of small grains to
the standard dust-to-gas mass ratio. The 
dust-to-gas mass ratio of the large grains
in the midplane was increased to conserve
the standard mass ratio at each radius \citep{d'alessio06}. The wall emission
was calculated following
%\tableddcomments{The wall's radius ({R}$\rm _{wall}$) and height (z$\rm _{wall}$) are calculated based on the best fit following
\citet{d'alessio05}. 
The maximum
grain size in the wall at the dust
destruction radius, as well as the dust
sublimation temperature, $\rm T_{wall}$,
and the height of the wall, $\rm z_{wall}$, were
kept as free parameters.

%We considered combinations of parameters using a grid of disk models. The range of values explored are shown in Table 2. The inclination, $\rm i$, and $\rm {R}_{d}$ of the disk were initially constrained using the ALMA observations (Section \ref{obs}). 
%Additionally, f
Following the results of the parametric model of the surface brightness profile (Section \ref{obs}), we also
included
a 
cavity in the 
%\textbf{mm} 
distribution of large dust grains 
%was included 
in some models. 
Here, we define ``cavity'' as a region
of the disk in which dust is significantly
depleted.
We note that the near-IR and mid-IR parts of the SED, which traced the small dust grains located close to the star, did not show significant depletion. Therefore, we included an inner cavity in our model by decreasing the abundance of only the large grains out to $\rm {R}_{cav}$ by a factor $\rm \delta_{cav}$. 
%We also varied \textbf{the maximum size of three grain-size distributions for grains in different parts of /the disk. These included the small grains in the wall, $\rm {a}_{max}$, small grains in the disk's atmosphere, $\rm {a}_{max}^{atm}$, and the large grains in the disk's midplane, $\rm {a}_{max}^{mid}$.} In addition, we adjusted the dust sublimation temperature of the wall, \textbf{$\rm T_{wall}$}, and the degree of dust settling for the disk, $\rm \epsilon$ \textbf{, which is determined from the relative mass abundance ratio of small grains to large grains}.

We considered combinations of parameters using a grid of disk models. The range of values explored are shown in Table 2. The inclination, $\rm i$, and $\rm {R}_{d}$ of the disk were initially constrained using the ALMA observations (Section \ref{obs}).
From the grid of parameters tested, the models that best fit the SED were obtained by minimizing the ${\chi}^{2}$. A synthetic image of each model was computed. The images were then Fourier transformed and sampled with the uv coverage of the ALMA observations, obtaining the simulated visibilities of the model. From these visibilities, an image was obtained using the CLEAN task in CASA. The simulated visibilities and images were compared with the observed visibilities and the averaged radial profile of the disk, respectively.

    %\subsubsection{Modeling as Pre-Transitional Disk}

\iffalse
\begin{figure*}
    \begin{subfigure}[t]{1\linewidth}
        \centering    
     \includegraphics[width=16cm,height=16cm,keepaspectratio, trim={0cm 0cm 0cm 0cm},clip]{fig3a.pdf}
    \end{subfigure}
    \centering
    
     \begin{subfigure}[t]{1\linewidth}
        \centering    
     \includegraphics[width=16cm,height=16cm,keepaspectratio, trim={0cm 0cm 0cm 0cm},clip]{fig3b.pdf}
    \end{subfigure}
    \centering
\fi

\begin{figure*}
        \centering  
     $\begin{array}{c}
     \includegraphics[width=16cm,height=16cm,keepaspectratio, trim={0cm 0cm 0cm 0cm},clip]{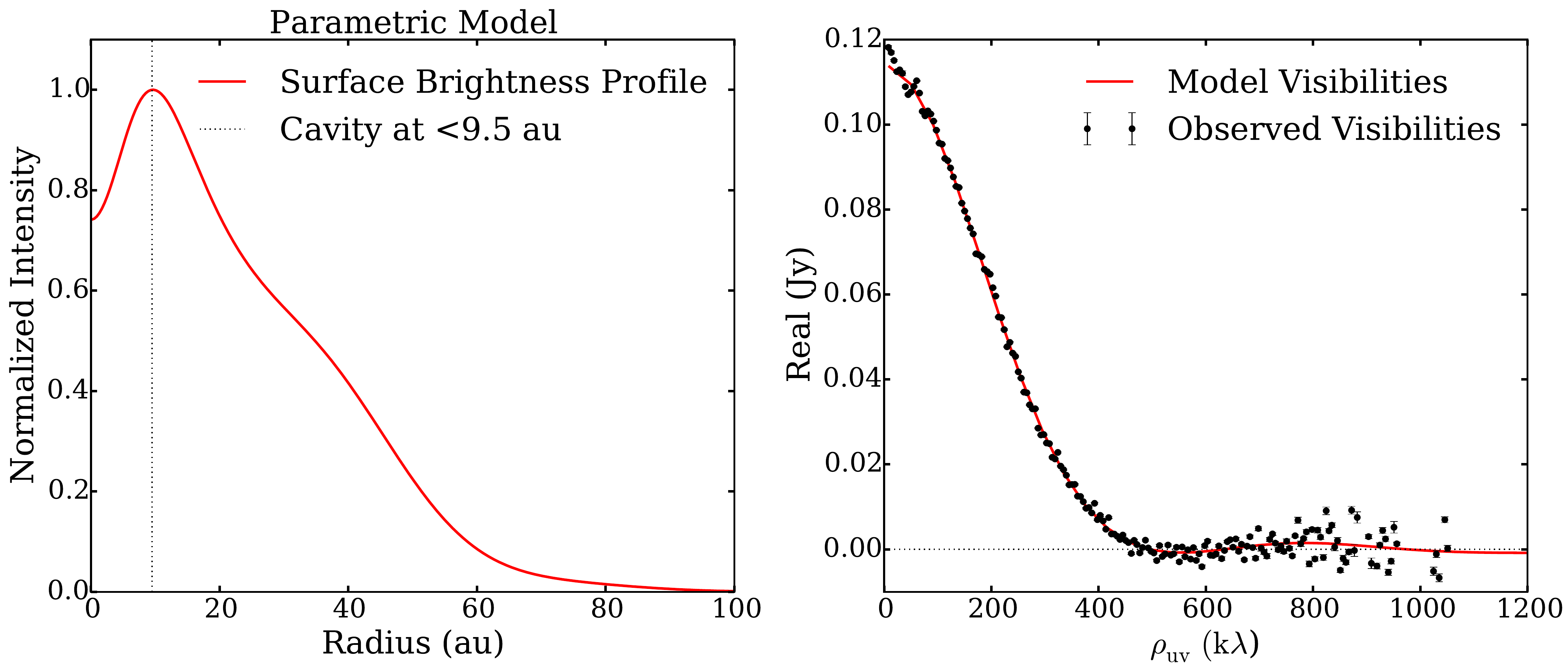}
     \\
     
     \includegraphics[width=16cm,height=16cm,keepaspectratio, trim={0cm 0cm 0cm 0cm},clip]{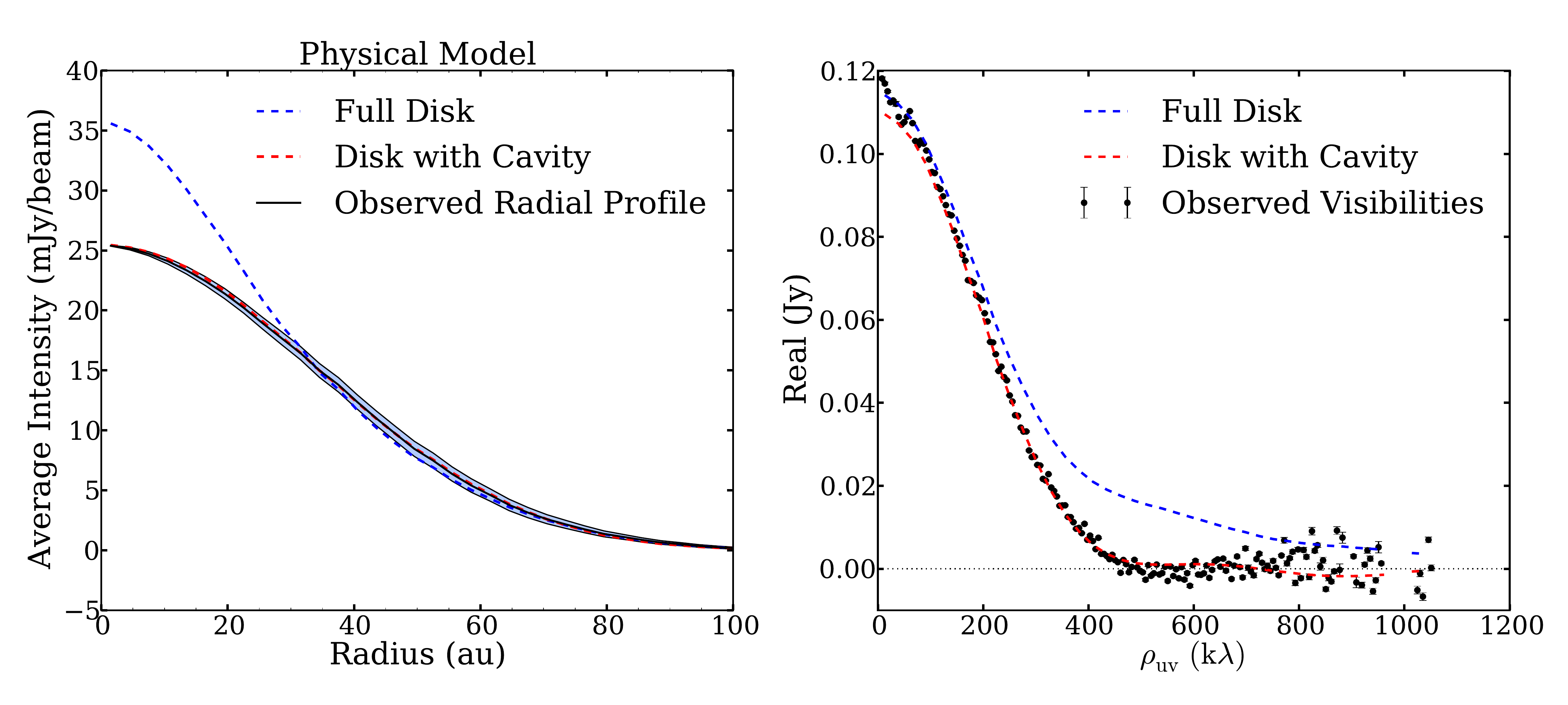}
     \\
     \end{array}$

    \caption{
    %Results from our analytical model of the surface brightness profile (\textit{Top}), and our physical disk model (\textit{Bottom}). The \textit{left} panels show the observed visibilities (black dots) and the fitted models (dashed curves). The \textit{right} panels show the intensity radial profiles of the best-fit models.
    \textit{Top} -- Results from our parametric model of the surface brightness profile of the disk. The normalized surface brightness profile of the best-fit model is shown on the left. The vertical line indicates the edge of a shallow cavity at \textless9.5 au. The right panel shows the observed (black dots) and model (red line) visibilities. \textit{Bottom} -- Results from our physical disk model. On the left panel, we show the averaged radial intensity profile of the observed image (black line), and of the simulated ALMA image obtained from the disk model (dashed lines). The error bars for the observed radial profile are computed based on the standard deviation of the disk emission at each radius. We note that the inner cavity is not clearly seen in the radial profile because of the angular resolution of the observations ($\sim$ 30 au). The observed (black dots) and model (dashed lines) visibilities are plotted on the right. The blue and red lines indicate the best-fit models using a full disk and a disk with a cavity of large grains, respectively. Based on the visibilities and the radial intensity profiles, the observations can only be fit when including an inner cavity in the disk.}
    \label{fig:fig_finalrad}
\end{figure*}

    \subsection{Disk Properties} 
    \label{res}
Figure \ref{fig_res} shows the final fit to the SED, and Figure \ref{fig_img_obs} shows the model (middle) and residual (right) images. The fit to the real part of the visibilities and to the radial profiles are shown in 
the lower panels of
Figure \ref{fig:fig_finalrad}. 
The parameters 
%used to fit 
that best fit
the SED and  
ALMA observations are shown in Table \ref{tab3}. 

Full disk models always failed to reproduce the visibilities and the inner parts of the radial profile (Figures \ref{fig_img_obs} and \ref{fig:fig_finalrad}). We found our best fit to the ALMA observations when including a cavity in the large dust grains out to 15.6 au. Nevertheless, the near-IR and mid-IR range of the SED fit well with a continuous abundance of small grains, suggesting that small dust grains are filling the cavity.
%an inner wall was needed to reproduce the NIR range of the SED, implying that HD 142666 is probably a pre-transitional disk. 

Our best fit also corresponded to 
an inner edge of the disk (R$_{
\rm wall}$)
of 1.3 au 
(Table \ref{tab3}). In comparison, 
\citet{schegerer13} and \citet{vural14}, 
obtained inner disk radii of
0.8 au and 1.35 au, respectively. 
These studies found additional structure
inside these radii using
near-IR and mid-IR visibilities. We do not
attempt to include these constraints
in this paper. However, 
since
near-IR emission arises in 
small dust in the atmospheric layers of the
inner disk 
%and the disk edge
\citep{d'alessio06}, the
\citet{schegerer13} and \citet{vural14}
results
are consistent with having small dust
down to scales $\sim$1 au, in agreement
with our results.

To find the dust mass of the disk, we integrated the surface density 
distribution of our best fit
%\textbf{From our 
disk model and found a mass of 5.3 $\times$ 10$^{-4}$ \Msun. Assuming a dust-to-gas mass ratio of 0.01, we then found a total disk mass of 0.0533 \Msun \, and a mass of 1.55 $\times$ 10$^{-7}$ \Msun \, for small grains in the disk's cavity. We also found a high degree of settling in the disk ($\rm \epsilon$ = 0.001), which was consistent with the classification of HD 142666 as a Group II source. As indicated by the ALMA observations, the disk of HD 142666 is also fairly compact, with an $\rm R_{d}$ of 65 au and a relatively sharp edge.
%The best fit used the inclination of $\sim$60$^\circ$ estimated from the ALMA observations, which was also consistent with the literature (Sect. \ref{obs}). 
%Finally, our best-fit \textbf{$\rm T_{wall}$} of 1100 K and \textbf{wall} radius of 1.3 AU \textbf{were} consistent with a near-IR interferometric study of the dust sublimation temperature for pre-main sequence star disks \citep{monnier05}. 

%\textbf{Finally, our best fit corresponded to a 
%wall radius of 1.3 au. 
%Previous works that estimated the wall radius, 
%including \citet{schegerer13} and \citet{vural14}, 
%found small dust grains outwards of at least $\sim$ 1 au 
%(1.35 au and 0.8 au, respectively).
%For comparison, scaling our wall radius 
%from 150 pc based on GAIA
%to a parallax distance of $\sim$115 pc 
%as used by these previous studies, 
%our wall radius is then $\sim$1 au. 
%While we did not fit the NIR and MIR visibilities as these works have,
%our wall radius is consistent with past studies in that no large cavity
%exists in the small grain distribution.}

%We note that we are unable to accurately fit the IRS spectrum of HD 142666 since it shows the presence of complex dust components, such as PAHs and crystalline silicates \citep[e.g.][]{mcclure10}, which we have not included in our model.

%%% TABLE 3:
%\begin{wraptable}{l}{1cm}
\begin{deluxetable}{ccc}

%\tabletypesize{\scriptsize}

\tablecaption{Model Properties}
\tablecolumns{3}
\tablewidth{0pt}
\tablehead{
\colhead{Parameter} & \colhead{Range of Parameters Tested} & \colhead{Best Fit}
}
\startdata
\tableline
\colhead{Wall} \\
\tableline

$\rm {a}_{max}$ (\mic) & 0.1 -- 100 & 1\\
$\rm {R}_{wall}$ (au)   & \nodata  &  1.3 \\
$\rm T_{wall}$ (K) & 1000 -- 1500 & 1100 \\ %, step $\Delta$log($a_{maxw}$)=1 & 1000.0\\ %JH: this was fitted?
$\rm {z}_{wall}$ (au) & 0.27 -- 0.4 & 0.32  \\
\tableline
\colhead{Cavity} \\
\tableline
$\rm {\delta}_{cav}$ & 1 -- 0.001 & 0.11\\
%$\rm \gamma$ & 1 -- 10 & 8\\ %, step $\Delta$log($a_{maxw}$)=1 & 1000.0\\ %JH: this was fitted?
%${R}_{cav, in}$ (au) & 1 -- 6 & 5\\
$\rm {R}_{cav}$ (au) & 8 -- 20 & 15.6\\
\tableline
\colhead{Disk} \\
\tableline
$\rm {a}_{max}^{atm}$ (\mic) & 0.1 -- 100 & 2\\
$\rm {a}_{max}^{mid}$ (cm) & 0.1 -- 1 & 1\\
$\rm \alpha$ & 0.01 & 0.01 \\
$\rm \epsilon$ & 0.001 -- 1 & 0.001 \\
$\rm i$ ($^{\circ}$) & 40 -- 70 & 60 \\
\Mdot \, (\msunyr) & 1 x ${10}^{-8}$ -- 2 x ${10}^{-7}$ & 7.5 x ${10}^{-8}$ \\
R$\rm _{d}$ (au) & 30 -- 100 & 65 \\
$\rm \gamma$ & 1 -- 10 & 8\\ %, step $\Delta$log($a_{maxw}$)=1 & 1000.0\\ %JH: this was fitted?

\enddata
\tablecomments
{The radius of the
wall, R$\rm _{wall}$, follows
from the temperature of the
wall, T$\rm _{wall}$,
%) and height (z$\rm _{wall}$) are calculated based on the best fit following 
following \citet{d'alessio05}. 
%$\rm \alpha$, the viscosity parameter \citep{shakura&sunyaev73}, was fixed because of its degeneracy with \Mdot. 
}
%\tablenotetext{a}{We refine our best fit  $\alpha$}
%\tablenotetext{d}{This value correspond to R$_{wall}$=9.5 AU}
\label{tab3}

\end{deluxetable}
%\end{wraptable}

%\twocolumn
\section{Discussion}
\label{discussion}  
From the results of our physical irradiated accretion disk model, we have inferred the presence of a cavity in large dust grains out to 15.6 au in the disk of HD 142666. These results are supported by the parametric model, based on the surface brightness profile, that fits the deprojected visibilities and indicates a significant decrease of mm emission at R\textless9.5 au. Our results are the first evidence of the presence of a wider cavity of large grains in the disk, which may have been elusive in the past because it is present only in the large dust grain distribution traced by the ALMA observations. Group II sources have been proposed to have their outer disks self-shadowed by the inner regions of the disk, which could result in a colder, ring-like region of the disk. Nevertheless, this cannot explain the cavity revealed by the ALMA observations since the radio emission probes dust in the midplane, where a shadow would have a very small effect. 

Several mechanisms have been proposed to explain the presence of cavities in protoplanetary disks \citep{espaillat14}. One mechanism involves dynamical interactions due to planets in the disk, which have been shown to open cavities and gaps in the dust distribution. Additionally, the edges of these cavities represent local pressure maxima, which can trap large dust grains, while allowing small dust grains to move towards the star and fill the cavity \citep{zhu14}. Thus, since the cavity in HD 142666's disk is present mainly in the large dust grain distribution, this cavity aligns with a planetary origin.

We compare the dust distribution
of HD 142666
with that of the Group I object
HD 100546.
%, which is representative
%of the class. 
ALMA observations
reveal $\sim$10 au cavities
in the large grains in both
objects \citep[this work; HD 100546, ][]{pineda14}. 
The largest difference is
in the distribution of
small grains. In
HD 100546, 
small grains populate 
the disk at radii $<$1 au,
producing near-IR excess.
%this small disk is
%found in other Group I objects
%\citep[e.g.][]{honda15}.
%small grains populate 
%the disk at radii $<$1 au,
%providing some near-IR excess.
Small grains also coexist with large
grains in the outer disk.
But a large region, 
between $\sim$1 au and 
$\sim$10 au, is essentially
depopulated of small grains \citep[based on AMBER/MIDI;][]{benisty10}.
%Small inner disk and
%large gaps in small grains
%are
%found in other Group I objects
%\citep[e.g.][]{honda15}.
In contrast, there is no
such large gap in the small grains in
the disk of
%the Group II object 
HD 142666;
%. The  small grains 
they are present in the disk
down to $\sim$1 au \citep[based on AMBER/MIDI;][]{schegerer13, vural14}.

%\textbf{Next, we 
%compare the distribution of small and large grains for the disks around HD 142666, a Group II source, and HD 100546, a Group I source which has similar observations and a potential protoplanet forming in its disk %\citep{walsh14}. 
%Beginning in the inner regions of the disk, HD 142666's disk contains a \textless 1 au gap in the small dust grain distribution starting at sub-au radii \citep[based on AMBER/MIDI;][]{schegerer13, vural14} and a \textless 15.6 au cavity in the large grain distribution (based on ALMA imaging, this work). In comparison, HD 100546's disk contains a much larger \textless13 au gap in the small grain distribution 
%\citep[based on H- and K-band AMBER/MIDI interferometry;][]{benisty10}, while the large dust grain distribution reveals an inner gap from the star out to a radius of 0.25 au and a gap at radii between 1 and 13 au \citep[also based on ALMA;][]{pineda14}. 
%Regarding the outer regions of these disks, HD 142666's disk contains a distribution of large and small grains that extend out to a similar, compact radius of $\sim$65 au \citep[respectively, ALMA in this work, scattered light with SPHERE in][]{garufi17}. While the large dust grain distribution of HD 100546's disk also extends out to a more compact radius of \textless 60 au \citep[based on ALMA;][]{pineda14}, the small dust grain distribution is more extended out to a radius of \textless546 au \citep[based on a distance from GAIA of $\sim$109 pc and SPHERE;][]{garufi16}}
    
More generally, prior studies in the near-IR to mid-IR showed that most Group I souces have large, \textgreater 20 au, inner clearings of dust \citep{maaskant13, honda15}, while cavities found in Group II sources have been small, \textless 1 au, and detected only close to the sublimation radius \citep{menu15}. Together with the proposed structural differences between the disks of Group I and Group II sources as indicated by their SEDs, these different cavities and structures have been explained with two scenarios: different evolutionary stages (a common ancestry) or separate evolutionary paths \citep{maaskant13, menu15, garufi17}. \citet{garufi17} postulated that as gaps are only present in the disks of Group I sources, the groups were more likely to evolve along separate paths.

Our results have showed that HD 142666 has an inner cavity extending out to $\sim$15.6 au. In addition, the Group II source HD 163296 was recently shown to have multiple gaps, as well as a possible shallow, inner gap with a width of $\sim$20 au \citep{muroarena18, zhang16, isella16}. 
Although the classification of HD163296 as Group II has been
disputed by \citet{muroarena18}, our 
%These 
results indicate that the characterization of  
%that 
Group I and Group II sources 
%have, respectively, 
as gapped and continuous disks may not always apply.

The inner cavities and gaps in the disks of Group II sources, as with those of HD 142666 and HD 163296, may be smaller and/or shallower than those in the disks of Group I sources, which may also explain why inner cavities and gaps of disks around Group II sources are not evident in SEDs. In fact, the SEDs of other Group II sources set constraints on the size of inner cavities, if present. Thus, even though cavities and gaps can be present in the disks of both Group I and Group II sources, there seems to be differences in the size of cavities and gaps between the the two groups. These differences cannot be explained by a scenario where Group I sources evolve into Group II sources, since inner gaps or cavities are not expected to decrease in size or to be refilled with dust during their evolution. An inverse process could be postulated, where the gaps of Group II sources get larger with time so that Group II evolves into Group I \citep[e.g.,][]{menu15}. However, as pointed out by \citet{garufi17}, the fact that disks around Group II sources have lower millimeter fluxes and smaller disk sizes than those of Group I sources does not favor this possibility.

Therefore, despite having shown that gaps can be present in disks around both Group I and Group II sources, different evolutionary tracks still seem to be the most likely scenario to explain both groups. The different gap sizes between Group I and Group II could in fact be a direct consequence of these different evolutionary tracks. If Group I includes denser and more massive disks, then they could be more likely to form giant planets. If so, such planets would open wider and deeper cavities than those around Group II sources, which, if present, would be formed by less massive planets.

In conclusion, our results, together with previous observations of HD 163296, show that the disks of Group II sources can contain cavities or gaps of large grains, 
despite prior claims. 
These inner gaps or cavities of large grains
may be,
%are, 
however, smaller than those seen around Group I sources. Our detailed modeling of HD 142666 also
indicates that no corresponding
cavities exist in the small grains. Higher angular resolution observations with ALMA will be necessary to confirm whether cavities are present in the disks of other Group II sources. Such studies will inform how disk structure relates to planet formation.

\section{Acknowledgments} 

We thank Megan Reiter, Thanawuth Thanathibodee, and Sierra Grant for useful discussions. A.E.R is grateful for the support given by the National Science Foundation Research Experiences for Undergraduates program through NSF grant 13-542. C.C.E acknowledges support from the Sloan foundation. K.Z. acknowledges the support of NASA through Hubble Fellowship grant HST HF2-51401.001-A awarded by the Space Telescope Science Institute, which is operated by the Association of Universities for Research in Astronomy, Inc., for NASA, under contact NAS-26555.

This research made use of the SIMBAD database, operated at the CDS, Strasbourg, France. This paper also makes use of the following ALMA data: ADS/JAO.ALMA\#2013.1.00498.S. ALMA is a partnership of ESO (representing its member states), NSF (USA) and NINS (Japan), together with NRC (Canada) and NSC and ASIAA (Taiwan) and KASI (Republic of Korea), in cooperation with the Republic of Chile. The Joint ALMA Observatory is operated by ESO, AUI/NRAO and NAOJ. The National Radio Astronomy Observatory is a facility of the National Science Foundation operated under cooperative agreement by Associated Universities, Inc.

%\section{References}
%\clearpage

%\bibliographystyle{apj}
%\section{References}
%\bibitem[Furlan et al.(2006)]{Furlan2006} Furlan, E., Hartmann, 
%L., Calvet, N., et al.\ 2006, \apjs, 165, 568

%% Connor's way:
%\bibliographystyle{apalike}
%\bibliography{biblio}

%\section{Appendix}\label{sect: Appendix}

\end{document}